\documentclass[12pt]{article}
\usepackage{amsmath}
\usepackage{amsthm}
\usepackage{esvect}
\usepackage[toc,page]{appendix}
\usepackage{leftidx}
\usepackage{color}
\usepackage{framed, color}
\usepackage{multirow}
\usepackage{pdfpages} 
\usepackage{multicol}
\usepackage{wrapfig,lipsum,booktabs}

\usepackage[utf8]{inputenc}
\usepackage{mathtools,hyperref} 
\usepackage{cleveref}
\usepackage{commath}

\usepackage{enumitem}
\usepackage{amssymb}

\usepackage{mathtools,hyperref}
\hypersetup{
    colorlinks=true,
    linkcolor=cyan,
    filecolor=cyan,      
    urlcolor=red,
    citecolor=red,
}
\usepackage[linesnumbered,ruled]{algorithm2e}

\definecolor{mgreen}{RGB}{25,147,100}
\definecolor{shadecolor}{rgb}{1,.8,.1}
\definecolor{shadecolor2}{RGB}{245,237,0}
\definecolor{orange}{RGB}{255,137,20}
\definecolor{orange}{RGB}{245,37,100}

\usepackage[capposition=bottom]{floatrow}
\usepackage{capt-of}
\usepackage{booktabs}
\usepackage{varwidth}
\usepackage[T1]{fontenc}
\usepackage[font=small,labelfont=bf,tableposition=top]{caption}
\DeclareCaptionLabelFormat{andtable}{#1~#2  \&  \tablename~\thetable}

\usepackage{mdframed}
\usepackage{adjustbox}
\usepackage{tcolorbox}
\usepackage{graphics}
\usepackage{tikz,ifthen,fp,calc} 

\usepackage{subcaption}
\usetikzlibrary{plotmarks}
\usepackage{graphicx}
\usepackage{capt-of}
\usepackage{booktabs}
\usepackage{varwidth}

\theoremstyle{plain}

\newtheorem{prop}{Proposition}[section]

\theoremstyle{definition}
\newtheorem{definition}{Definition}[section]

\theoremstyle{definition}
\newtheorem{remark}{Remark}[section]

\usepackage[english]{babel}
\usepackage{babel,blindtext}

\usepackage{fullpage}
\usepackage{amsfonts}
\usepackage{lscape}
\usepackage{bbm}

\usepackage{todonotes}
\usepackage{cite}
\usepackage{verbatim}
\usepackage{bm}

\usepackage[numbers]{natbib}

\usepackage{authblk}

\usepackage[margin=1in]{geometry}
\title{Loss of community identity in opinion dynamics models 
as a function of inter-group interaction strength}

\author[1]{Hossein Noorazar \thanks{h.noorazar@wsu.edu}}
\author[1]{Matthew J. Sottile\thanks{mjsottile@math.wsu.edu}}
\author[1]{Kevin R. Vixie\thanks{vixie@speakeasy.net}}
\affil[1]{Department of Mathematics and Statistics, Washington State University}

\providecommand{\keywords}[1]{\textbf{\textit{Keywords:---}} #1}

\usepackage[utf8]{inputenc}
\date{}
\begin{document} 
\date{}
\maketitle
\date{}
\begin{abstract}
Recent technological changes have increased connectivity between
individuals around the world leading to higher frequency interactions
between members of communities that would be otherwise distant and
disconnected.  This paper examines a model of opinion dynamics in
interacting communities and studies how increasing interaction
frequency affects the ability for communities to retain distinct
identities versus falling into consensus or polarized states in which
community identity is lost.  We also study the effect (if any) of
opinion noise related to a tendency for individuals to assert their
individuality in homogenous populations.  Our work builds on a model
we developed previously~\cite{Noorazar2016} where the dynamics of
opinion change is based on individual interactions that seek to
minimize some energy potential based on the differences between
opinions across the population.
\end{abstract}

\\
\section{Introduction} \label{sec:introduction}
During the past several years, scientists have built up different models
 to explain dynamics of opinion evolution in a society~\cite{Proskurnikov2018, Proskurnikov2018a}.
 They are called opinion models and have been used 
 to study the dynamics of a set of individuals who 
 communicate their opinion on one or more topics 
 and adjust them based on these interactions.

Early models typically reach a steady state within 
a small number of interactions (e.g. \cite{DeGroot1974,FRENCH1956}) such that the population 
reaches consensus, polarization, or a state in which a set
of distinct opinions are held indefinitely by subpopulations. 
Subpopulations with similar opinions are referred to as clusters.

 In the case in which a set of subpopulations 
 exist where each subpopulation reaches a 
 distinct equilibrium opinion state, it is useful to study
the effect of parameters on the system that affect 
communities' ability to maintain these distinct 
opinion states such that they do not collapse together.  
Collapse implies that these subpopulations are no longer
distinguishable by having distinct opinion sets.  In this work, we
consider the consensus equilibrium opinion states to be the \emph{identity} of
each population. These identities reflect shared opinions on any
topic, be it volatile topics (such as politics or religion), or benign
topics (such as locally popular makes and models of cars).

The stability of identity of such subpopulations is of relevance in
social systems where subpopulations correspond to meaningful groups,
such as families, townships, political groups, religious groups, and
so on.  Under what conditions can such subpopulations coexist and
maintain their distinct identities with some degree of cross-group
interaction?  What factors cause them to either merge to a consensus
state or diverge to a state in which they adopt polar opposite
opinions on a given topic?  This is particularly relevant to study in
the current world in which subpopulations of people that traditionally
have coexisted and maintained different, but non-polarizing opinions,
may find it difficult to maintain these distinct identities with
respect to specific topics.  In particular, the emergence of
high-frequency, high-reach communication mechanisms that did not exist
prior to the 21st century (e.g., social media) fundamentally change
the interaction dynamics between individuals and communities compared
to lower frequency, lower reach mechanisms in the past.  

The variety of factors which play different 
roles in real social life where the invention of 
Internet and technology has brought people
together and closer than ever, make the dynamics
complex. 
Therefore, to keep it simple we study two such factors.
We would like to investigate:
\begin{itemize}
\item Stability of communities as a function of 
inter-community interaction rate.

\item Effect of  symmetric and asymmetric noise on stability of communities
in presence of inter-community interactions.
\end{itemize}

\subsection{Opinion models}

For the purpose of formal modeling and analysis, we adopt an idealized
numerical model of opinions.  An opinion is considered to be a
numerical value within a range (e.g., $[0,1]$) called opinion space.  Individuals are
simple agents that hold opinion values on one or more distinct topics.
Agents interact in a pairwise fashion to exchange and update their
opinion state over time.  In the model that we adopt for this work
(detailed in~\cite{Noorazar2016}), individuals update their opinions after
interactions in a manner that reduces the energy of some potential
function.  A brief summary of the relevant components of the model for
this paper are described in Section~\ref{sec:model}.

\subsection{Stability and robustness of subpopulations}

\begin{figure}[httb!]
\centering
\begin{subfigure}{.3\textwidth}
  \includegraphics[width=1\linewidth]{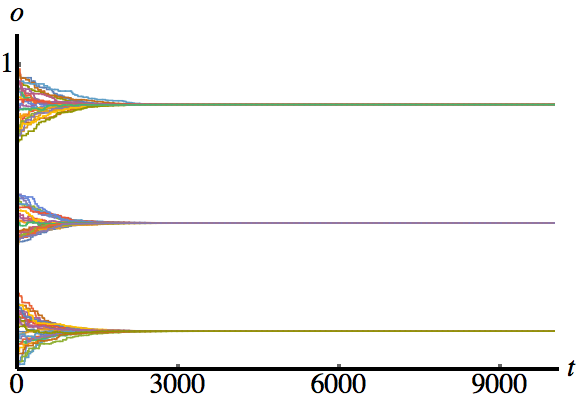}
  \caption{}
  \label{fig:noInterCommunityInteractionFull}
\end{subfigure}
\hspace{.05in}
\centering
\begin{subfigure}{.3\textwidth}
  \includegraphics[width=1\linewidth]{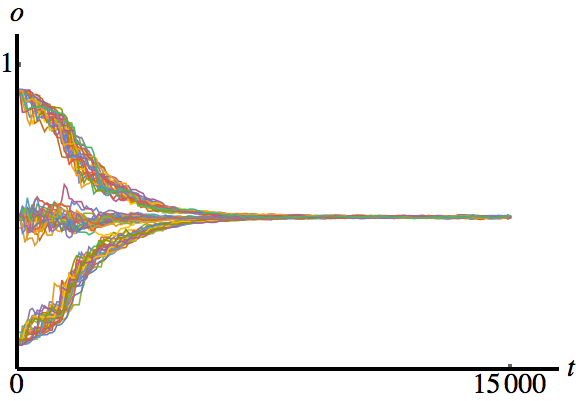}
  \caption{}
  \label{fig:Lostidentity1}
\end{subfigure} 
\hspace{.05in}
\centering
\begin{subfigure}{.3\textwidth}
  \includegraphics[width=1\linewidth]{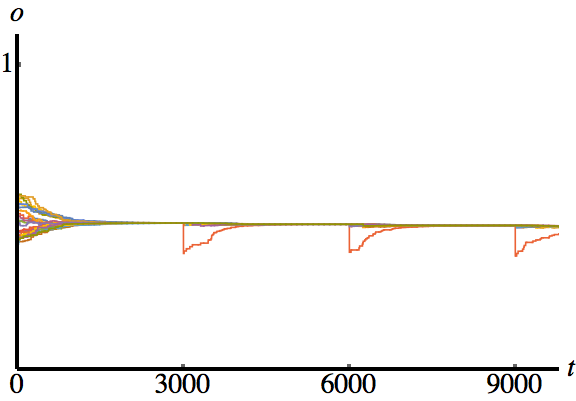}
  \caption{}
  \label{fig:PerturbationFull}
\end{subfigure} 
\hspace{.05in}
\centering
\begin{subfigure}{.3\textwidth}
  \includegraphics[width=1\linewidth]{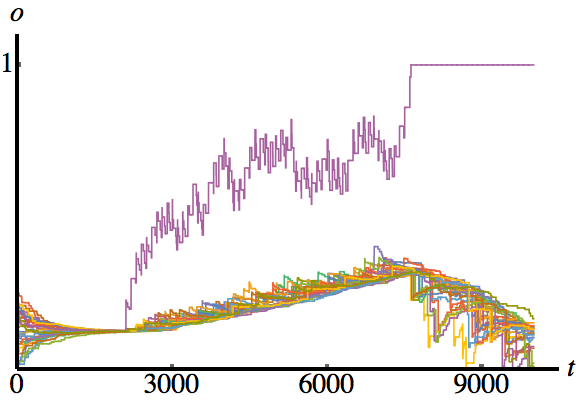}
  \caption{}
  \label{fig:frequentPerturbation}
\end{subfigure}%
\hspace{.05in}
\centering
\begin{subfigure}{.3\textwidth}
  \includegraphics[width=1\linewidth]{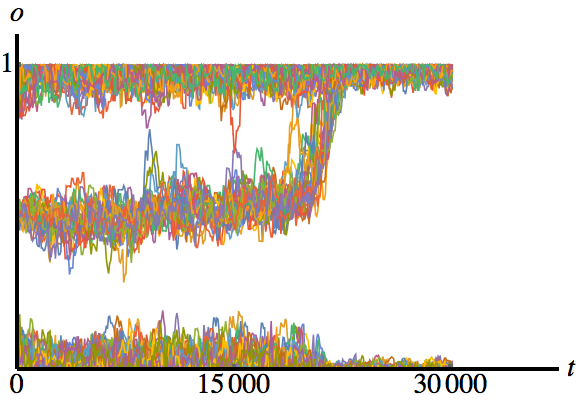}
  \caption{}
  \label{fig:Lostidentity}
\end{subfigure}%
\caption{\textbf{Stable communities, abrupt changes and inter-community interactions.}
\small \tiny In part~\ref{fig:noInterCommunityInteractionFull} there 
are 3 communities with no inter-community interaction and they all
go to their equilibrium states. 
In part~\ref{fig:Lostidentity1} there are high inter-community 
interactions and the three communities come to consensus 
and become stable. Part~\ref{fig:PerturbationFull}
includes one agent who makes abrupt changes and gets pulled back 
to the community it belongs to with minor change of equilibrium state
of the group. 
However, if a given agent changes its opinion
too frequently, it could cause instability of
the community~\ref{fig:frequentPerturbation}.
 (please, note that one time step here means one single pairwise interaction.)
}
\label{fig:StableCommunitiesPerturbation}
\end{figure}

In this work we will consider a single large population of individuals
in which a set of subpopulations exist, each of which contains a
subset of the population whose collective opinion state is in
equilibrium.  An equilibrium opinion state is reached when
interactions between members of the subpopulation do not reduce the
overall interaction energy (e.g., the individuals are in a consensus
state, (Figs.~\ref{fig:noInterCommunityInteractionFull} and \ref{fig:Lostidentity1}\footnote{Agents in Fig.~\ref{fig:Lostidentity1} 
follow the bounded confidence model rule.}).  
We consider a stable subpopulation to be one in which the
introduction of a new opinion by one individual that is sufficiently small
and  different from the equilibrium state of the 
subpopulation will not cause the overall opinion 
of the population to change by more than some $\epsilon$.
After introduction, the individual holding the 
new opinion will eventually be driven towards 
the equilibrium state of the subpopulation 
such that the resulting new equilibrium state 
will be within $\epsilon$ of the state before the 
new opinion arrived. We would say that this 
deviation from the stable point will be damped out
within a finite number of interactions
(Fig.~\ref{fig:PerturbationFull}).  If this deviant opinion retains
its difference from the stable state of the subpopulation from which
it emerged, such as by additional deviation before it can be damped
out, we could see a difficulty in the subpopulation to damp the
opinion out (Fig.~\ref{fig:frequentPerturbation}) leading to an overall
effect on the equilibrium state of the subpopulation.

A reinforcement effect emerges when more than one member of the
subpopulation makes opinion changes at a frequency above the time
scale by which the population can damp out the deviations and return
to their previous stable state.  When such events occur, we expect
that previously stable subpopulations may destabilize and fracture,
driving them to new equilibrium states. Under an interaction potential
function that supports multiple stable minima (such as a potential
supporting both consensus and polarization states), we may observe a
rapid disintegration of the subpopulation at equilibrium into a set of
polarized subpopulations due to a frequency of introduction of
divergent opinions that exceeds the timescale necessary to damp out
deviations from equilibrium.  When this threshold is passed 
we would see a loss of identity for subpopulations.
They may be driven to a polarized state in which individuals adopt opinions
at the extrema of the opinion space (Fig.~\ref{fig:Lostidentity}), or a homogenous consensus state in
which no distinct subpopulations exist anymore (Fig.~\ref{fig:Lostidentity1}).

Our hypothesis is that this frequency of deviation is the cause of
fracturing of otherwise stable communities into those in which
homogenous consensus or polarized opinion states dominate.  The
interesting result is that populations that would otherwise agree may
rapidly split due to high-frequency introduction of opinions outside
their consensus state, which is precisely what emerges in social
networks where interaction frequency with large populations outside
one's own community is commonplace.

\subsection{Individualization tendency}

In real life people would like to be unique and different from others,
as such our model includes opinion noise representing this natural desire.

\begin{figure}[httb!]
\centering
\begin{subfigure}{.35\textwidth}
  \includegraphics[width=1\linewidth]{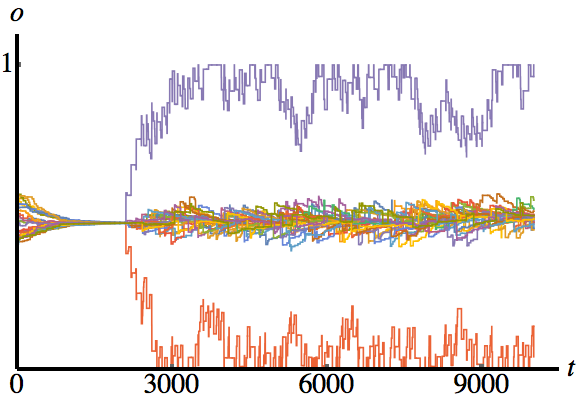}
  \caption{}
  \label{fig:oppositeDirecFrequent}
\end{subfigure}%
\hspace{.2in}
\begin{subfigure}{.35\textwidth}
  \includegraphics[width=1\linewidth]{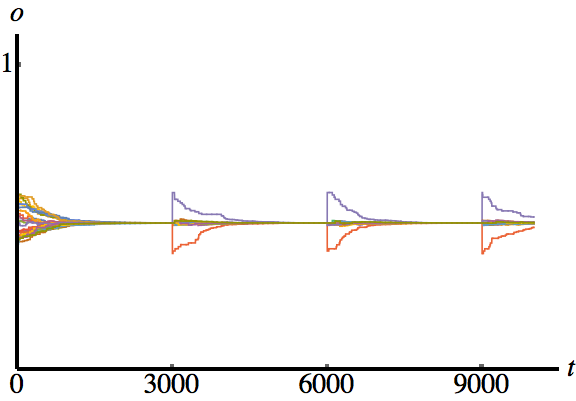}
  \caption{}
  \label{fig:oppositeDirectioninFrequent}
\end{subfigure}%
\caption{\textbf{Opposite abrupt changes and its frequency.} 
}
\label{fig:Oppositeperturbationanditsfrequency}
\end{figure}

It is shown that opinion noise is an important contributing factor to
the existence of opinion clusters~\cite{Mas2010}.  In this paper we
also study its effect on the stability of interacting subcommunities
and whether there exists a relationship between the frequency of
interactions and properties such as bias or magnitude of noise.
Noise can have different effects in terms of the response of a subcommunity
to the introduction of new opinions away from equilibrium.  Noise may cause
an individual to counterbalance the effect of the new opinion, providing
an opposing force that cancels out the deviation.  Similarly, noise may
reinforce the deviation if it is in the same direction, compounding its
effect. 
In Fig.~\ref{fig:Oppositeperturbationanditsfrequency} we see two agents
decide to be different from the rest of the group.  In
Fig.~\ref{fig:oppositeDirecFrequent} they go their way too
frequently and cause the group to have a wider range of opinion about
previous equilibrium state that in a long run could divide the group
into two smaller groups.  But as can be seen in
Fig.~\ref{fig:oppositeDirectioninFrequent} if the abrupt change is
not too frequent, the community could save its identity.

We study two types of noise: 
noise in which the change in opinion is symmetric
about the equilibrium state, and noise where an individual is more likely to
move their opinion in the opposite direction of their subpopulation peers.

\subsection{Concepts in the paper} \label{sec:concepts}
\subsubsection{Damping perturbations from equilibrium}

Communities do not lose 
their identity when a single member makes a
single (sufficiently small) abrupt change ($\Delta$) in their opinion on a given topic.
This is due to a damping process that occurs where subsequent repeated
interactions within the community with the individual who made the
change gradually pulls that individual back to the consensus opinion
held by that community.  This process is not instantaneous, as each
interaction changes the opinions of individuals by a small amount
$\delta << \Delta$. If $\Delta \approx n\delta$, and we expect one
interaction with the individual who made the large jump every $t$ time
units, then we would say that the deviation of size $\Delta$ has a
damping period of $nt$ over which time the subpopulation absorbs the
deviation and recovers its collective identity with respect to the
given topic.

The frequency of deviations is critical.  If another deviation occurs
before the damping period has elapsed, the deviations may reinforce
each other (if they went in the same direction relative to the
consensus state) and make it take longer for the entire subpopulation
to return to its collective consensus state.  When opinion change of
nodes are due to interactions across subpopulations, then we must
consider the relationship of the damping time necessary to recover
from an individual deviation and the frequency of interaction between
subpopulations that cause deviations to occur.


\subsubsection{Frequency of inter-group interaction}
Probability of interaction between two nodes is used
as a surrogate for frequency of interaction between them. (See Section~\ref{sec:NetWorkConnectovity}). These probabilities 
are stored in adjacency matrix as edge weights between nodes.
Higher probability of interactions causes more frequent interactions.
\subsubsection{Noise via individualization tendency}

A somewhat different type of opinion change, 
\textit{an abrupt change}, is introduced to 
the system by adding the individualization tendency to the model. 
The individualization tendency we use here is an \textit{``adaptive''} one, i.e.
agents desire to leave their community increases according
to the two following factors: 
(a) the difference between a given agent's opinion and other agents'
opinion decreases and, 
(b) number of agents whose opinions are close to a 
given agent increases. In other words, let opinion of agent $i$ 
be given by $o$, and let $(o-\epsilon, o+\epsilon)$ be a 
neighborhood of $o$, then individualization tendency of agent $i$ 
becomes more intense as the number of other agents 
in this neighborhood increases or as $\epsilon$ decreases.
This is given by Eq.~(\ref{eq:adaptive-noise}). 
This noise is drawn from a normal distribution 
with a zero mean and adaptive variance. 

\subsection{Contributions}
The model we present in this paper makes some
 noteworthy contributions to the study of opinion dynamic.
We conduct computer experiments to show that:
 \begin{itemize}
 \item Increased frequency of interaction
 between subcommunities causes identity loss.

\item Individualization tendency makes 
the identity loss of communities to take place 
more often and faster.
\end{itemize}
\section{Model} \label{sec:model}
First let us start with some definitions and notation that will
be used later in the paper.

\theoremstyle{definition}
\begin{definition}{}
Let $G = ( V , E ) $ represent the \emph{fully connected network} under consideration
where $V$ is the set of nodes and $E$ is the set of edges in which
all distinct nodes are connected via an edge with a probability of interaction weight
assigned to each edge. 
The set of all \emph{spatial neighbors} of node $i \in V$ is denoted by $N(i)$
contains all nodes in $G$ that are connected to $i$ via an edge
with an interaction probability weight of more than zero.
\end{definition}

\theoremstyle{definition}
\begin{definition}{}
The set of all possible (numerical) opinions, denoted by $\mathcal{O}$, 
is called \textit{opinion space}, and in this paper it will be the interval [0,1].
\end{definition}

\theoremstyle{definition}
\begin{definition}{}
A \emph{social group} is the set of all nodes in $G$
with high probability of interactions. These nodes
are located on diagonal blocks of the adjacency matrix of $G$.
\end{definition}

\theoremstyle{definition}
\begin{definition}{}
A \emph{opinion cluster} or a \emph{(opinion) community} is the set of all nodes
which agree about a topic. In other words, set of all nodes
whose opinion belong to $(o - \epsilon, o + \epsilon )$ for some $\epsilon$. 
Opinion $o$ is called the \emph{identity} of such community.

\end{definition}

In the experiments, members of a given social group
hold the same opinion at time $t=0$.

\theoremstyle{definition}
\begin{definition}{}
Define the \emph{$\delta$-opinion-neighbor} of node $i$ 
at time $t$, denoted by $\mathcal{P}_\delta^t(i) $, 
 to be the set of all nodes whose opinion are in 
 $(o_i^{(t)} - \delta, o_i^{(t)} + \delta )$,  for some $\delta$ at time $t$ . 
\end{definition}

\theoremstyle{definition}
\begin{definition}{}
\emph{Individualization tendency}, $\xi$, is a noise randomly chosen
from a normal distribution with zero mean and some variance 
$\sigma$ $(\xi \sim N(0, \sigma))$.
\end{definition}

\begin{tcolorbox}
\begin{center}
\textbf{Notation used throughout the paper}
\end{center}
\begin{multicols}{2}
[
]
\begin{itemize}
\item Matrices will be shown by bold letters.
\item $\textbf{I}_k$ identity matrix of size $k$.
\item $\textbf{1}_k$ matrix of ones of size $k$.
\item $n_s$ number of social groups.
\item $n_p$ population of each social groups.
\item $N = n_s \times n_p$ total population of the network.
\item $N(i)$ set of spatial neighbors of agent $i$.
\item $\mathcal{O}$ opinion space.
\columnbreak
\item $o_i^{(t)}$ opinion of node $i$ at time $t$.
\item $\alpha$ learning rate.
\item $\psi$ potential function.
\item $\xi_i(t)$ agent $i$'s noise at time $t$.
\item $N(\mu, \sigma)$ normal distribution with mean $\mu$ and standard deviation $\sigma$.
\item $s_p$ skewness parameter.
\item $N(s_p, \mu, \sigma)$ skew normal distribution 
with skewness parameter $s_p$, mean $\mu$ and standard deviation $\sigma$.

\end{itemize}
\end{multicols}
\end{tcolorbox}
\subsection{Network connectivity}\label{sec:NetWorkConnectovity}
In this paper, we work with a fully connected graph in which
all individuals are connected via an edge with a 
weight assigned to it. The weight corresponds to the 
probability that agent $i$ will talk to $j$ in a single step.
The network adjacency matrix \textbf{A} is a block symmetric doubly-stochastic 
matrix whose entries, $a_{ij}$, determine the probability 
of agent $i$ choosing agent $j$ for  an interaction, i.e. weights assigned to
edges in $G$ are stored in \textbf{A}.
Lets denote the number of social groups by 
$n_s$ where each of them have equal population of size $n_p$. 
Then the matrix \textbf{A} would be of the size $ N \times N$
where $N = n_p \times n_s $ is the population of the network.
\[ 
\textbf{A} =
\begin{bmatrix}
    \textbf{A}_{11}  & \textbf{A}_{12}  & \dots  & \textbf{A}_{1,n_s} \\
    \textbf{A}_{21}  & \textbf{A}_{22}  & \dots  & \textbf{A}_{2,n_s} \\
    \vdots & \vdots & \ddots & \vdots \\
    \textbf{A}_{n_s,1} & \textbf{A}_{n_s,2}  & \dots  & \textbf{A}_{n_s,n_s}
\end{bmatrix}
\]

The entries in diagonal blocks $\textbf{A}_{kk}$, $1 \le k \le n_s$ 
are interaction probabilities of agents within a social group while
 the entries in $\textbf{A}_{lk}$, $1 \le l \neq k \le n_s$ are  
 probabilities of inter-group interactions.
 
 The adjacency matrix will be generated in two different ways,
deterministically and randomly. Consequently two different 
 sets of experiments and analysis will be represented.

\subsubsection{Deterministic adjacency matrix}\label{MPSection:AdjacencyDet}

In the first case scenario, the adjacency matrix is generated deterministically by 
Algorithm~\ref{alg:DetadjacencyGenAlg} in which the probability 
of interaction between members of a social group are identical,
and the interaction probability between any two agents of different
groups are the same.

\begin{algorithm}[httb!]
\tiny
    \SetKwInOut{Input}{Input}
    \SetKwInOut{Output}{Output}
    \Input{Two nonnegative integers $n_s$, $n_p$ and $i_p \in [0,1]$}
    \Output{deterministic adjacency matrix \textbf{A}}
    $N = n_s \times n_p$;                     \hspace{.1in} //population size\\
    \textbf{A} = zero matrix of size $N$    \hspace{.1in}  // initiate adjacency matrix\\
    // Initiate diagonal blocks:\\
    \For{$k = 1, \ldots, n_s$ }{%
       $a^k_{ij} = 1/(n_p-1)$ \hspace{.1in} if $ i \neq j $
      }
       $\textbf{A}_{ij}  \gets i_p \times \textbf{1}_{n_p} - \textbf{I}_{n_p}$  // generate off-diagonal blocks\\ 
       $\textbf{A} \gets \textbf{A}/sum(\textbf{A}[1,:])$// make it doubly stochastic\\
       return \textbf{A}
    \caption{\small Deterministic adjacency matrix generation.(DAMG)}
    \label{alg:DetadjacencyGenAlg}
\end{algorithm}

\subsubsection{Random adjacency matrix}\label{MPSection:AdjacencyMG}
In this method entries of off-diagonal blocks are randomly chosen
from an interval bounded above by some \emph{upper bound} $u_b \in [0,\: 1]$. 
 So, when $u_b = 0$ then there would not be any
 interaction between two different communities. 
 We use the iterative method defined by Sinkhorn 
 \cite{Sinkhorn1964} to generate the doubly-stochastic
 matrices with a little modification to make 
 them symmetric (Algorithm \ref{alg:adjacencyGenAlg}).

At the end of this process, the entries of 
diagonal blocks $\textbf{A}_{kk}$ will not be equal, 
which means that, each pair of nodes in the same 
social group will not have the same frequency, 
i.e. probability, of interaction.

 Note that $\textbf{A}_{ij}$ is a submatrix and 
 $a_{ij}$ is a single entry of \textbf{A} and we use $a^k_{ij}$ 
to denote a single entry of the block $\textbf{A}_{kk}$.

\begin{algorithm}[httb!]
\tiny
    \SetKwInOut{Input}{Input}
    \SetKwInOut{Output}{Output}
    \Input{Two nonnegative integers $n_s$, $n_p$ and $u_b \in [0,1]$}
    \Output{adjacency matrix \textbf{A}}
    $N = n_s \times n_p$;                     \hspace{.1in} //population size\\
    \textbf{A} = zero matrix of size $N$    \hspace{.1in}  // initiate adjacency matrix\\
    // Initiate diagonal blocks:\\
    \For{$k = 1, \ldots, n_s$ }{%
       $a^k_{ij} = 1/(n_p-1)$ \hspace{.1in} if $ i \neq j $
      }
       // generate off-diagonal blocks:\\ 
       \For{$i = 2, \ldots, n_s$ }{%
          \For{$j = 1, \ldots, i-1$ }
          {%
          $\textbf{A}_{ij}$  choose randomly from $[0,u_b)$ \\
          $\textbf{A}_{ji} \gets \textbf{A}_{ij}$ // copy lower diagonal blocks to upper diagonal blocks\\
           }
      }
      \While{not converged}
      {
      \For{$rowCount = 1, \ldots, \text{N}$}
      {%
      // divide each row by sum of its entries:\\
      $\textbf{A}[rowCount,:] \gets \textbf{A}[rowCount,:]/sum(\textbf{A}[rowCount,:])$
      }%

      \For{$colCount = 1, \ldots, \text{N}$}
      {%
      // divide each column by sum of its entries:\\
      $\textbf{A}[:, \: colCount ] \gets \textbf{A}[:,colCount]/sum(\textbf{A}[:,colCount])$\
      }%
            \For{$i = 2, \ldots, n_s$ }{%
          \For{$j = 1, \ldots, i$ }
          {%
          $\textbf{A}_{ji} \gets \textbf{A}_{ij}$ // copy lower diagonal blocks to upper diagonal blocks\\
           }
      }
      } 
      return \textbf{A}
    \caption{\small Random adjacency matrix generation.(RAMG)}
    \label{alg:adjacencyGenAlg}
\end{algorithm}

\begin{figure}[httb!]
\centering
\begin{subfigure}{.2\textwidth}
  \includegraphics[width=1\linewidth]{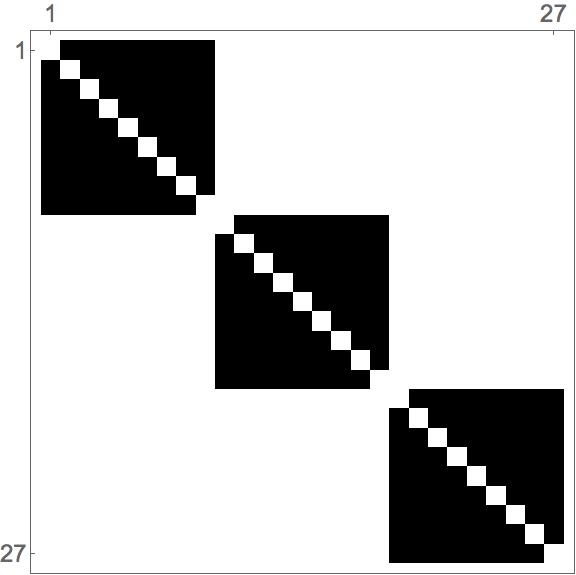}
  \caption{$u_b = 0$}
  \label{fig:adj-heat-ubZero}
\end{subfigure}%
\hspace{.2in}
\begin{subfigure}{.2\textwidth}
  \includegraphics[width=1\linewidth]{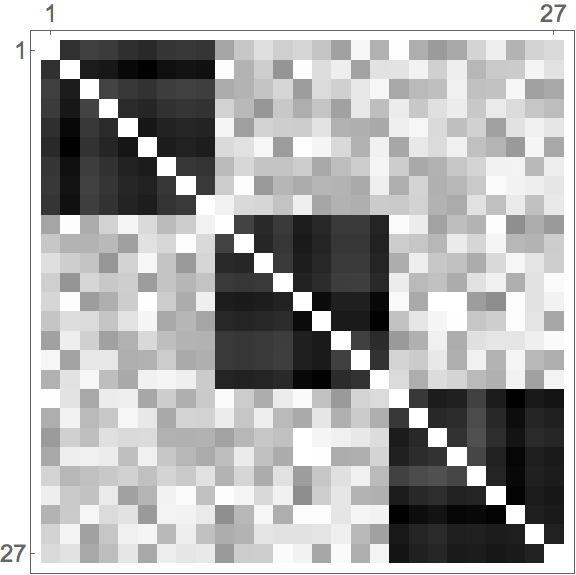}
  \caption{$u_b = 1/16$}
  \label{fig:adj-heat-ubOne}
\end{subfigure}%
\hspace{.2in}
\begin{subfigure}{.2\textwidth}
  \includegraphics[width=1\linewidth]{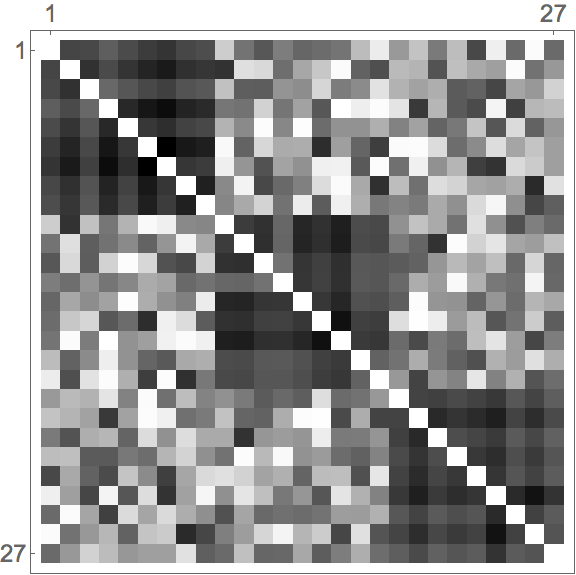}
   \caption{$u_b = 1/8$}
  \label{fig:adj-heat-ubEqual}
\end{subfigure}%
\hspace{.2in}
\begin{subfigure}{.2\textwidth}
  \includegraphics[width=1\linewidth]{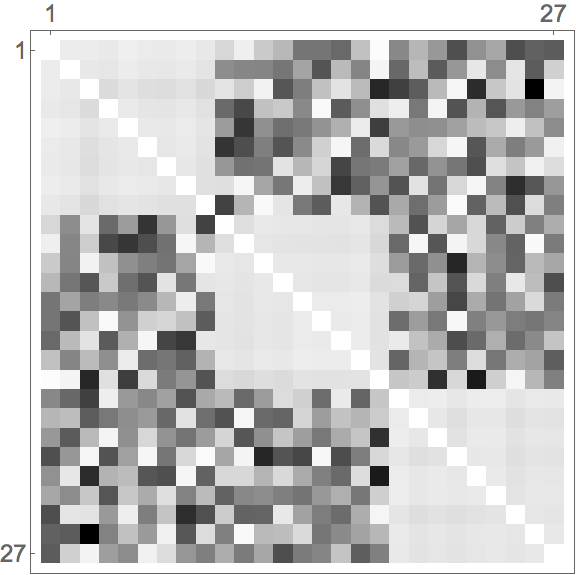}
   \caption{$u_b = 1$}
  \label{fig:adj-heat-ubOne}
\end{subfigure}%
\caption{\textbf{Adjacency matrix. }
\tiny There are three subcommunities, 
each subcommunity has a population of 9.
As $u_b$ is increases frequency of inter-communities
interaction increases and when $u_b=1$, (\ref{fig:adj-heat-ubOne}),
 subcommunities basically do not exist.
}
\label{fig:AdjacencyHeat}
\end{figure}

Figure~\ref{fig:AdjacencyHeat} is an (heat map) example of 
adjacency matrix with $n_s = 3$ groups, each group
 is consist of $n_p = 9$ nodes. 
 As $u_b$ increases the frequency of inter-group
 interaction goes up in expense of local group interactions
 to the point that inter-group connections are tighter than
 group connections ($u_b = 1$). When $u_b$ is larger than $(n_p - 1) ^{-1}$, 
it is possible that some entries in the off-diagonal blocks  
to be larger than the entries within the diagonal blocks. 
 
In (\ref{fig:adj-heat-ubEqual}) the upper bound 
is set to be $1/(n_p-1) = 1/8$, and we see there are places 
where frequency of inter-group interactions are higher 
than that of internal ones. 
The internal group interactions are diluted. 
The reason that we can still see subpopulations are 
relatively knitted tightly is that the entries of off-diagonal 
blocks are chosen randomly from $[0, \: 1/8)$
to begin with. So, lots of them start with values smaller 
than 1/8.
We generated 1000 of such matrices for $n_s = 3$, 
$n_p = 9$, and $u_b = 1/8$. 
Therefore, 3000 blocks of size 9 of subgroups. 
The mean and standard deviation of entries of
diagonal blocks, mostly belong to the intervals 
$(0.056, 0.062)$ and $(0.004, 0.008)$, respectively.

\subsection{Micro dynamics} \label{sec:updateRules}

Pairwise interaction rule of the model is borrowed from~\cite{Noorazar2016}.
Let opinion of agent $i$ at time $t$ be given by $o_i^{(t)} \in \mathcal{O}$. 
Then the update rule is given by:

\begin{equation}
\left\{
  \begin{array}{lr}
    o_i^{(t+1)}  &= o_i^{(t)}  - \frac{\alpha}{2} \:  \psi'(|d^{(t)}_{ij}|) \; d^{(t)}_{ij} + \xi_i(t)    \vspace{.1in}\\
    o_j^{(t+1)} &= o_j^{(t)}  +  \frac{\alpha}{2} \: \psi'(|d^{(t)}_{ij}|) \; d^{(t)}_{ij} + \xi_j{(t)}
  \end{array}
\right.
\label{eq:updateRule}
\end{equation}
where $\alpha$ is called learning rate, 
$\psi$ is called potential function which governs
the update rule, and  $d_{ij}^{(t)} = o_i^{(t)} - o_j^{(t)}$. 
If boundary condition of opinion space is violated,
then opinions will be clamped.
$\xi_i(t) \sim N(0, \sigma_i(t))$ adapted from~\cite{Mas2010},
is individualization tendency of agent $i$ at time $t$. It incorporates the 
natural instinct of people wanting to be different
~\cite{SnyderPursuit, Imhoff2008, Hornsey2004}.
$\xi_i(t)$ is randomly sampled from a normal distribution 
$ N(0,\sigma_i(t))$ where $\sigma_i(t)$ is
given by: 
 (The opinion space in \cite{Mas2010} is $[-250, 250]$.
So, we had to scale the variance to fit our opinion space
which is [0,1].)

\begin{equation}
\label{eq:adaptive-noise}
\sigma_i(t) = \frac{s}{e-1} \left( - |N(i)| + \sum_{j \in N(i)} e^{1-|d_{ij}(t)|} \right)
\end{equation}
where $N(i)$ is set of (spatial) neighbors of node $i$,
 the parameter $s$ is used to manipulate the strength 
of individualization tendency. 
Individualization tendency increases 
when there is high uniformity.
The individualization tendency mentioned above
does not take into account direction of movement of a given opinion cluster.
It is equally probable that an individual makes an 
abrupt change in any direction. 
Therefore, we also consider  a system 
in which agents have a memory in the sense that
they will consider direction of movement of the opinion-cluster
they belong to, so that it is more probable for them to move in 
the opposite direction of cluster's movement.

In order to increase probability of individual tendencies to be in the opposite direction of
cluster movements, we sample $\xi_i(t)$ from a skew normal distribution 
( $\xi_i(t) \sim N(s_p, 0, \sigma_i(t))$), defined by \cite{Azzalini}, 
where $s_p$ is \emph{skewness parameter} and variance is defined as before.
The skew parameter of such a distribution is given by $s_p = - c_m s_s$, where $s_s$ 
is a constant called \emph{skewness strength}, and $c_m$, defined below determines 
the direction of movement of the cluster to which agent $i$ belongs to.

\theoremstyle{definition}
\begin{definition}{}
Let $\mathcal{P}_{\delta}^t(i)$, be the set of $\delta$-opinion-neighbor
of agent $i$ at time $t$ for some $\delta$. Then define $c_m$
to be the direction that majority of $\mathcal{P}_{\delta}^t(i)$ moves towards ($sgn(x)$ is the sign function):

\[c_m = sgn \left[ \sum_{j\in \mathcal{P}_{\delta}^t(i)} sgn \left( o_j^{(t)} - o_j^{(t-1)} \right) \right] \]
\end{definition}
\subsection{Equilibrium response to abrupt individual changes}

Let $\hat \Delta := \Delta_i^{(t)}$ be the deviation 
from consensus made by individual $i$ on the given 
topic at time $t$.  During subsequent 
interactions between $i$ and members 
$j \in \mathcal{P}_0^t(i) \setminus \{i\}$, we expect 
that $o_i$ will move closer to $o_{\mathcal{P}_0^t(i) \setminus \{i\}}$.
\begin{prop}\label{prop:driftProp}
Let $\mathcal{P}_0^0(i) = \{i, i_2, i_3,  \ldots , i_{k} \}$ and 
suppose there is no inter-community interaction
and let the opinion of group to be $\hat o \equiv o_{\mathcal{P}_0^0(i)}$ 
at $t=0$. 
Let the potential function for all agents to be the same and
suppose agent $i$ makes an abrupt change: $o_i = \hat o + \hat \Delta \in \mathcal{O}$
(WLOG assume $\hat \Delta > 0$),
so that it is in attraction domain of the subgroup, then 
the equilibrium state of the members of subgroup will drift by $\hat \Delta /k$.
\begin{proof}
since $\hat \Delta$ is small enough
so that other agents in the group will attract agent $i$,
for all agents in $\mathcal{P}_0^0(i)$ define the energy $e_j$ 
to be the height from $\hat o$. Then total energy of
the given community is $E:= \hat \Delta$ at $t=0$.
Since all agents are using the same potential function,
the step size that a pair of node makes in a single interaction 
is the same. 
Therefore, for example, in the first interaction between $i$ and $i_j$
we have $o_i^{(1)} = \hat o + \hat \Delta - \epsilon_1$ and $o_{i_j}^{(1)} = \hat o + \epsilon_1$.
There will not be loss of energy. 
Hence, after $t = m $ steps, where $m$ is large enough, we must have 
$E = \hat \Delta + 0 + \ldots + 0 = \frac{\hat \Delta}{k} + \frac{\hat \Delta}{k} + \ldots + \frac{\hat \Delta}{k}$. 
Therefore, all agents will come to consensus at $\tilde{o} = \hat o + \frac{\hat \Delta}{k}$
and the drift size is $\frac{\hat \Delta}{k}$.
\end{proof}
\end{prop}
\begin{remark}[] 
Note that in the Prop.~\ref{prop:driftProp} it is assumed
that $\hat \Delta$ is sufficiently small so that agent $i$
will be attracted to the subgroup it belongs to. But, for example
if a tent potential given by Eq.~(\ref{eq:TentPotential}) is used
and $\hat \Delta > \tau$, then agent $i$ will repel the rest of the group
, agent $i$ and the subgroup will end up in opposite extreme points 
of opinion space. Or, if a bounded confidence potential is used,
where agents will not interact if they are far enough, then
after the abrupt change agent $i$ and the rest of the group
might ignore each other and everyone would stay where they are.
\end{remark}

\begin{remark}[] 
In any interaction, both nodes involved 
would take steps of the same size. 
Hence, average of opinions of all nodes 
has to be the same all the time. 
\[ \forall t: \frac{1}{N} \sum_i o_i^{(0)} = \frac{1}{N} \sum_i o_i^{(t)} \]

\noindent However, for example in Fig.~\ref{fig:Lostidentity} we see that
the average of all opinions at $t=30,000$ is 0.75 which is
different from average of opinions at $t=0$ which is 0.5.
The reason is that the boundaries of opinion space, does not let
the nodes to take steps of the right size and therefore,
system loses energy and average of opinion of all nodes
would not be the same.
\end{remark}

\section{Experiments}
Experiments are done to fairly small populations. 
They can be extrapolated to larger populations. 
However, there would be some differences. For example,
by~\ref{prop:driftProp} for a large population, the drift would
be smaller, therefore, longer time is needed for communities
to collapse. Or, for a fixed $i_p (u_b)$, as population $N$
increases, the probability of interaction inside a group
gets smaller and smaller relative to inter-group interaction
probabilities. Therefore, more time is needed to reach a steady state.

\subsection{Methods}
In the experiments\footnote{The codes for this experiments 
and that of ~\cite{Noorazar2016} can be found here: https://github.com/HNoorazar/} 
pairwise interactions follow the update rule given by Eq.~(\ref{eq:updateRule})
with learning rate $\alpha = 0.1$, and a tent potential function defined by Eq. (\ref{eq:TentPotential}) with $\tau = 0.63$. 
Initial opinion of communities are $0.1$, $0.5$ 
and $0.9$ respectively so that they are in an equilibrium state. 

\begin{equation}
\psi(x) = \begin{cases}  
\:  \: \: \frac{1}{\tau} x, &  x \leq \tau, \\
\frac{1}{\tau - 1 } x, & o.w.
\end{cases}
\label{eq:TentPotential}
\end{equation}

Individualization tendency will be sampled 
from a normal and skewed normal distribution with the 
same mean and variance.  Note that a time step is equal to $N$ pairwise interactions.

\subsection{Deterministic adjacency experiments}
\subsubsection{Population size and inter-group interaction rate}
Suppose there are $n_s$ subgroups in the network where each
group is consist of $n_p$ nodes. 
Then each column of the matrix has $N$ 
entries, of which $n_p = \frac{1}{n_s}N$ of them lie on a
diagonal block submatrix and $\frac{n_s-1}{n_s}N$ of them 
lie on the off-diagonal blocks. 

In the first step of generating the adjacency matrix,
$\frac{1}{n_p-1}$ is assigned to diagonal blocks,
and $i_p$ is assigned to off-diagonal blocks (Algorithm~\ref{alg:DetadjacencyGenAlg}),
denote such a matrix by $\textbf{A}_0$.
For example, in the case of $n_s=2$ and $n_p=2$ the adjacency matrix 
(at the beginning of the process) looks like:

\begin{equation}\label{eq:matrixExample}
\textbf{A}_0 =
\begin{bmatrix}
    0 & 1  & i_p  & i_p \\
    1 & 0  & i_p  & i_p \\
    i_p & i_p  & 0  & 1 \\
    i_p & i_p  & 1 & 0 \\
\end{bmatrix}
\end{equation}

In such a setting we are interested in knowing what 
is the relationship between the population size $N$ and
the smallest $i_p$ for which the social groups lose 
their identity and collapse. 
Hence, in order to find a relation between the two parameters
we do the following. For a given column of the matrix $\textbf{A}_0$, we look
at the ratio of sum of entries that lie on off-diagonal blocks to the 
sum of entries lying on diagonal blocks which is given by $ r_{i_p} = \frac{n_s-1}{n_s}Ni_p$.
For example, in the matrix given by Eq.~(\ref{eq:matrixExample})  
the ratio is $r_{i_p}=2i_p$.

 Let us denote the smallest $i_p$ that causes community 
 collapse by $i_{pc}$ and the corresponding ratio by $r_{i_{pc}}$. 

\begin{figure}[H]
\includegraphics[width=1\linewidth]{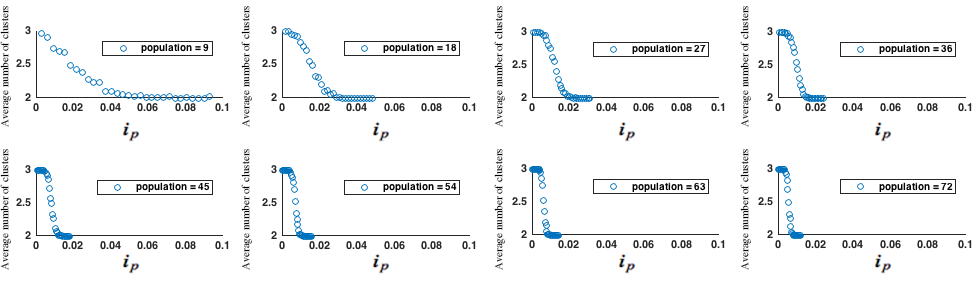}
\caption{\textbf{Minimum inter-group interaction probability to break communities.}}
\label{fig:Minupperb}
\end{figure}

Experiments with $n_s=3$, different $n_p$'s and $i_p$'s
are run and its results are shown by Fig.~\ref{fig:Minupperb} and 
Table~\ref{table:unifPotTable}. For most experiments, the ratio
is about $r_{i_{pc}}\approx0.34$.

\begin{table}
\begin{center}
    \begin{tabular}{| l | l | l | l | l | l| l| l| l | l | l | p{1cm} |}
    \hline
    $N$               & 9 & 18  & 27 & 36 & 45 & 54 & 63 & 72  \\ \hline
    \# of games   & 160 & 200 & 240 & 280 & 320 & 360 & 420 & 440 \\ \hline
\# of iterations  & 3,750 & 7,500 & 11,250 & 15,000 & 18,750 & 22,500 & 26,250 & 20,000 \\ \hline
$r_{i_{pc}}$ & 0.2790 & 0.3072 & 0.3420 & 0.3264 & 0.3420 & 0.3420 & 0.3570 & 0.3446\\ \hline
    \end{tabular}
\end{center}
\caption{Minimum inter-group interaction probability to break communities.}
 \label{table:unifPotTable}
\end{table}

\subsubsection{Individuality tendency causes identity loss}
\begin{figure}[!ht]
    \centering
      \includegraphics[width=.45\linewidth]{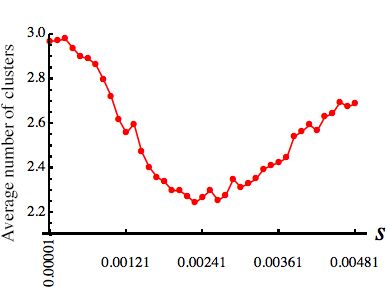}
\hspace{.2in}
\tiny
    \begin{tabular}[b]{l c c c}
       \multicolumn{2}{ c }{\tiny $ i_p = 0.0025$ }  \\ \hline
       \tiny $n_s$  & \tiny 3    \\ \hline
       \tiny $n_p$  & \tiny 4     \\ \hline
       \tiny $\Delta s$  & \tiny $12 \times 10^{-3}$ \\ \hline
       \tiny {\# of games per $s$}  & \tiny 500  \\ \hline
       \tiny \# of iterations per a game & \tiny 8000        \\ \hline
       \vspace{.4in}
    \end{tabular}
    \captionlistentry[table]{A table beside a figure}
    \captionsetup{labelformat=andtable}
    \caption{\textbf{Individuality tendency causes identity loss}($i_p=0.0025$).}
     \label{fig:sCauseLoss}
  \end{figure}

In order to see effect of individuality tendency on
identity loss, we pick up an small inter-group probability of
interaction, $i_p=0.0025$ and vary $s$. We can see in Fig.~\ref{fig:sCauseLoss}
that average of number of opinion clusters goes down as individuality tendency
goes up, and causes the communities to lose their identities.
However, after some point when $s$ grows, individuality tendency 
causes the social groups to earn a new identity and hence, we
see after $s =  0.0023$ where average number of clusters 
reaches its minimum at $2.24$, the average number of clusters
start to grow again. For example, in Fig.~\ref{fig:greatPoint} 
the parameters are $u_b = 0.01$ and $s = 0.0047$ 
(This figure has used random adjacency matrix). 
The plot shows how the individuality tendency causes communities 
lose their identities and then try to be different. The same result holds
for the random adjacency matrix as well.

\begin{figure}[httb!]
\includegraphics[width=.4\linewidth]{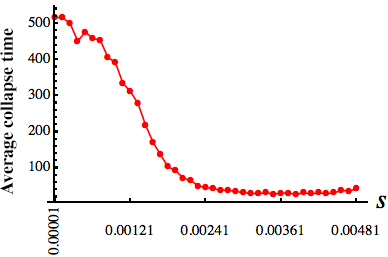}
\caption{\textbf{Individuality tendency accelerates identity loss ($i_p=0.0025$).}
\tiny Please note than the averages are taken over the cases in which communities have lost their identity. For example, according to Table 2, for a given pair of $(i_p, s)$ 500 games are done. If in 200 of them the communities lose their identity and in 300 do not, then 
the 300 games are not taken into account for computing the average collapse time.
Furthermore, the averages are taken for the smallest $t$ for which the communities collapse.
For example, in Fig.~\ref{fig:greatPoint} we see communities collapse early in the process,
but then again, they look for new identities to be different from a big group. }
\label{fig:tendEffectcollapseTimeDet}
\end{figure}

Figure~\ref{fig:tendEffectcollapseTimeDet} also shows, 
the more people want to be different, the sooner the 
communities will lose their identity. Please note that these
are averages of the times for which the communities lose their
identity first over time.

\subsection{Random adjacency experiments}
\subsubsection{No individualization tendency}
\label{sec:noNoiseSec}
\begin{figure}[httb!]
\centering
\begin{subfigure}{.35\textwidth}
  \includegraphics[width=1\linewidth]{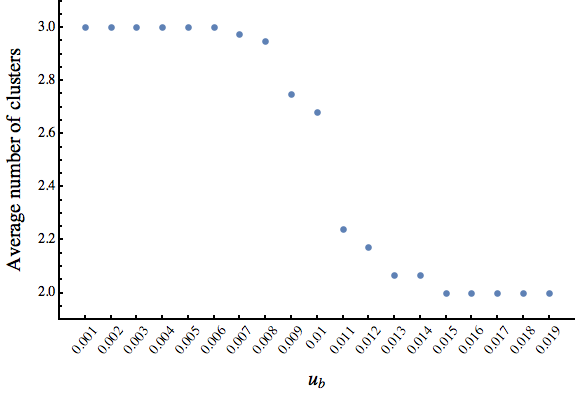}
  \caption{ Average number of clusters ($s=0$).}
  \label{fig:2dClusterUppEffectNoUniqNormal}
\end{subfigure}%
\hspace{1in}
\begin{subfigure}{.35\textwidth}
  \includegraphics[width=1\linewidth]{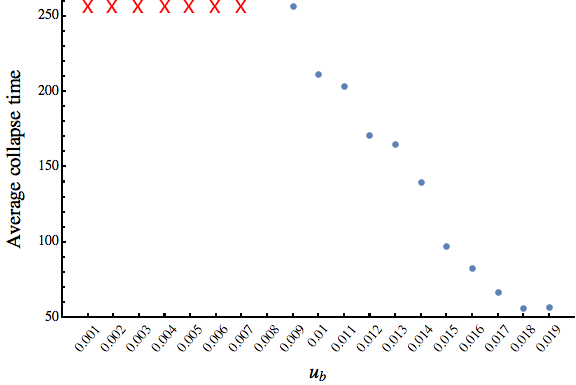}
  \caption{Average collapse time ($s=0$).}
  \label{fig:2dTimeUppEffectNoUniqNormal}
\end{subfigure}%
\caption{\textbf{Effect of inter-group interaction rate on community collapse.} 
\tiny There is no individualization tendency in this experiment ($s = 0$). In \ref{fig:2dClusterUppEffectNoUniqNormal} 
the dashed line shows that at $u_b = 0.01$ in almost half of experiments identity loss occurred.
In \ref{fig:2dTimeUppEffectNoUniqNormal} for those values of
$u_b$ which we have red {\color{red}{x}}, communities did not loose their identities
and did not collapse over the experiment run time (30,000 pairwise interactions).}
\label{fig:EffectOfinter-communityinteraction}
\end{figure}

First lets look at the case in which there is no individualization tendency, i.e. 
$s=0$, and $u_b$ is varied (Fig.~\ref{fig:2dClusterUppEffectNoUniqNormal}).
Having two clusters means communities have merged, 
since there was not an experiment in which
all three communities come to consensus, 
due to the choice of initialization of opinions and the fact
that the tent potential function has two minima.
We can see as frequency of interaction goes up, 
frequency of vanishing the community in the middle 
goes up as well and the middle community loses its 
identity and is combined with the other two. 

Furthermore, the average time needed for community
collapse has an inverse relation with inter-community
interaction rate. The more inter-community interaction frequency,
the less the time needed for communities to merge into one 
(Fig.~\ref{fig:2dTimeUppEffectNoUniqNormal}).

In some of the experiments, the community whose 
opinions are in $[ 0.49, 0.51] $ at $t = 0$ collapses 
and merges with the other two communities. 
 Average of collapsing time is taken just over the collapsed cases.

\subsubsection{Symmetric tendency of individualization}
\label{sec:symmetricNoiseSec}
\begin{figure}[httb!]
\centering
\begin{subfigure}{.3\textwidth}
  \includegraphics[width=1\linewidth]{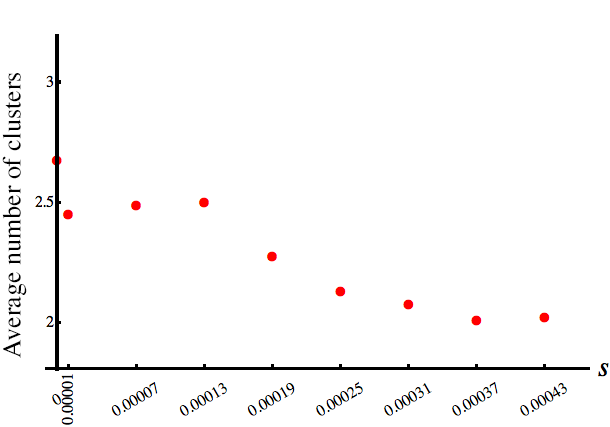}
\caption{Average number of clusters.}
\label{fig:2dClusterUniqEffectNormal}
\end{subfigure}%
\centering
\hspace{.01in}
\begin{subfigure}{.3\textwidth}
  \includegraphics[width=1\linewidth]{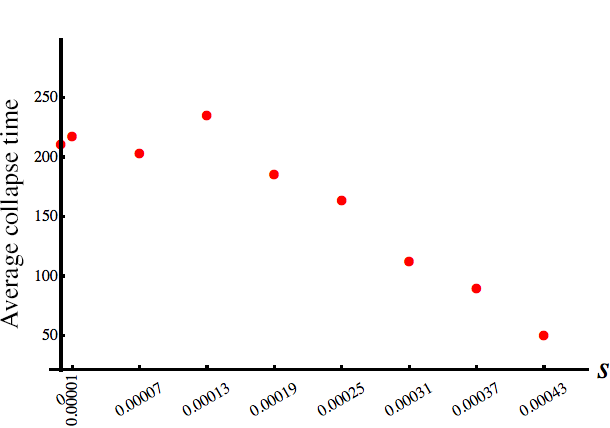}
  \caption{ Average collapse time.}
  \label{fig:2dTimeUniqEffectNormal}
\end{subfigure}%
\hspace{.01in}
\begin{subfigure}{.3\textwidth}
  \includegraphics[width=1\linewidth]{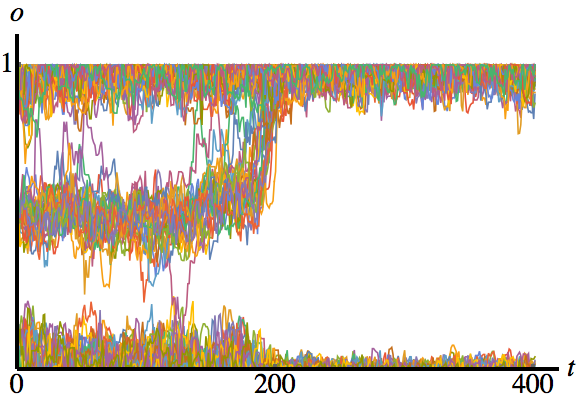}
  \caption{Not everyone can be unique! \\($s = 0.0043$).}
  \label{fig:fig5Bexplanation}
\end{subfigure}%
\caption{\textbf{Individualization effect on community collapse ($u_b = 0.01$).}}
\label{fig:TendencyForindividualizationEffect}
\end{figure}

In this section we sample individualization tendency
from a (symmetric) normal distribution. In this case
the noise can be in any direction and cluster's movement direction
is not taken into account. An individual can either make
an abrupt change in the direction that its neighbors are moving towards,
or in the opposite direction.
 In the experiment whose result is represented by 
Fig.~\ref{fig:2dClusterUppEffectNoUniqNormal}
we see that for $u_b = 0.01 $ 
we get approximately  $50\%$ of identity loss 
(i.e. $50\%$ of the experiments the community in the middle 
is merged with the other two and in $50\%$ of experiments it 
maintained its own identity). 
We set $u_b = 0.01$ and 
vary the individualization tendency strength.
The result is shown in  Fig.~\ref{fig:2dClusterUniqEffectNormal}.
Individualization tendency will cause the  identity loss to take 
place more frequently. 

Figure~\ref{fig:2dClusterUniqEffectNormal} and \ref{fig:2dTimeUniqEffectNormal}
have an interesting message. 
The more one wants to be unique, the less s/he can be.
Figure~\ref{fig:fig5Bexplanation} is illuminating. At time $t=0$
people in middle community have the urge to be different from 
other members of their own. They deviate themselves and get
close to the other two communities at the extreme points, 
and then interaction forces, bind them to the other communities.
Afterwards, any interaction they make is either from the members 
of their opinion-neighbors or they interact with agents in the other 
side of the boundary, which in both cases, they are forced 
to sit where they are!
Lets look at different combinations of $u_b$ and $s$.
Average number of clusters for each pairs are computed. 
We see as both $u_b$ and $s$ increase,
the community in the middle is effected by the other two and will
merge into them. And change of $u_b$ has more effective consequences.

\begin{figure}[httb!]
\centering
\begin{subfigure}{.45\textwidth}
  \includegraphics[width=1\linewidth]{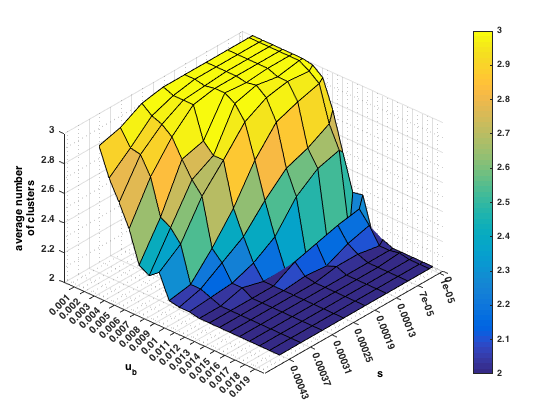}
  \caption{Average number of clusters.}
  \label{fig:cluster3DtentNormal}
\end{subfigure}%
\hspace{.1in}
\begin{subfigure}{.45\textwidth}
  \includegraphics[width=1\linewidth]{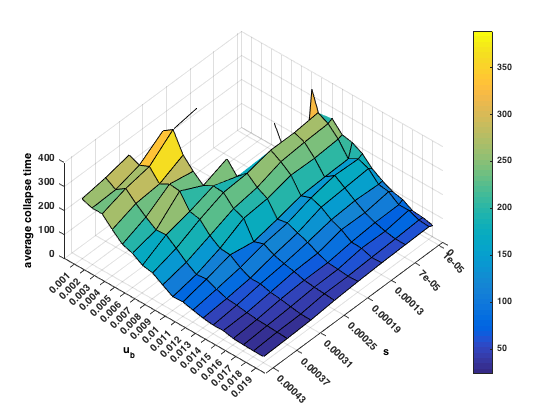}
  \caption{Average time of community collapse.}
  \label{fig:3dTimeNormal}
\end{subfigure}\\%
\vspace{1pt}
\caption{\textbf{Averages for different settings of $(u_b, s)$.}
\tiny Please note that in \ref{fig:3dTimeNormal} there
are missing pieces at the top right corner of the plot.
The missing pieces correspond to the cases
in which all three communities hold onto their 
opinions and did not merge into any other two,
to the final step of experiment.
}
\label{fig:IdoNotKnow}
\end{figure}

The missing pieces in Fig.~\ref{fig:3dTimeNormal} correspond
to the experiments in which all communities kept their identity. 
In Fig.~\ref{fig:2dTimeUniqEffectNormal}, which is 
one slice of Fig.~\ref{fig:3dTimeNormal}
we see that as individualization tendency and
collapse time have an inverse relation.

\subsubsection{Asymmetric individualization tendency}
\label{sec:AsymmetricNoiseSec}
\begin{figure}[httb!]
\centering
\begin{subfigure}{.35\textwidth}
  \includegraphics[width=1\linewidth]{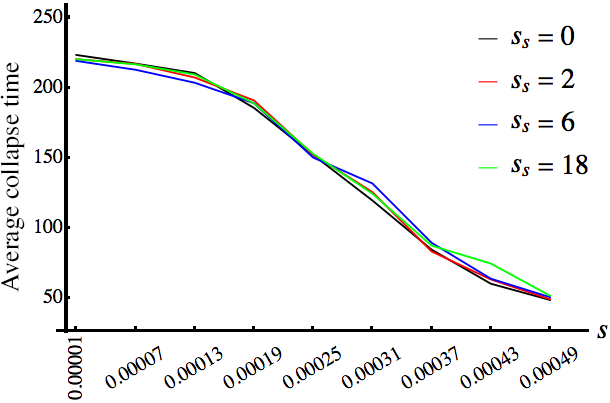}
  \caption{\tiny Random adjacency matrix ($u_b = 0.01$).}
  \label{fig:RandomAdjacencyMatrix}
\end{subfigure}%
\hspace{.3in}
\begin{subfigure}{.35\textwidth}
  \includegraphics[width=1\linewidth]{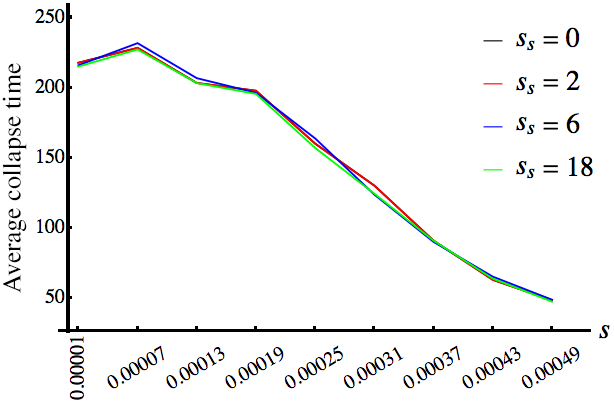}
  \caption{\tiny Deterministic adjacency matrix ($i_p = 0.005$).}
  \label{fig:DeterministicAdjacencyMatrix.}
\end{subfigure}\\%
\vspace{1pt}
\caption{\textbf{Average community collapse time.}}
\label{fig:AsymmetricIndividualizationTendency}
\end{figure}

In Fig.~\ref{fig:AsymmetricIndividualizationTendency}, 
where game is replicated 1000 times per $s$, 
the average time needed for community collapse 
for both deterministic and random adjacency matrices
are compared with four different skewness strength. 
Individualization tendency distributions have the same mean and
variance in all cases, however, the skewness parameters 
are different. In one case there is no skewness, i.e. 
distribution is symmetric, and three asymmetric
distributions of individualization samples 
with different skewness strengths are experimented so that
the individualization tendency is more probable to be
in the direction opposite to direction of movement of cluster
to which each agent belongs to.
It is interesting that the collapse times are almost identical
in almost all cases.

\section{Community matters}
\begin{figure}[httb!]
\centering
\begin{subfigure}{.3\textwidth}
\includegraphics[width=1\linewidth]{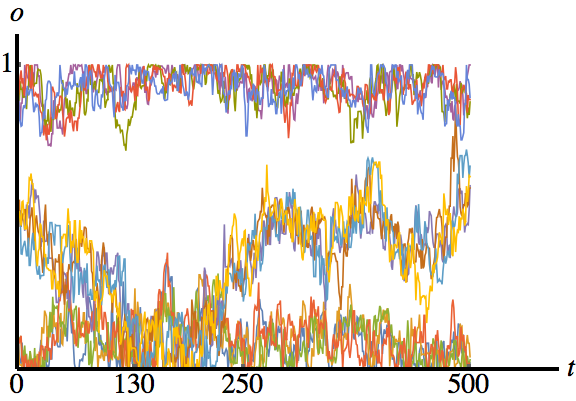}
 \caption{}
\label{fig:communitiesStickTogether}\end{subfigure}%
\centering
\hspace{.01in}
\begin{subfigure}{.3\textwidth}
\includegraphics[width=1\linewidth]{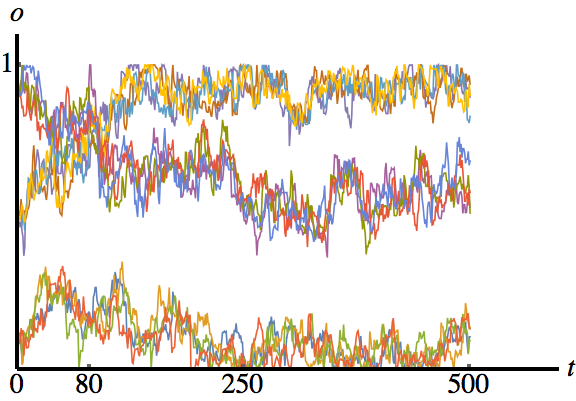}
  \caption{}
\label{fig:communitiesStickTogetherMatt}
\end{subfigure}%
\hspace{.01in}
\begin{subfigure}{.3\textwidth}
\includegraphics[width=1\linewidth]{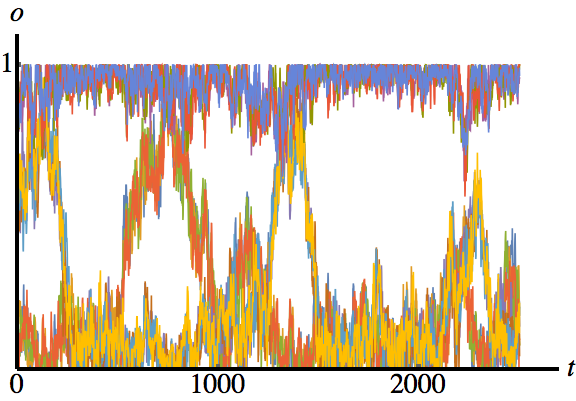}
  \caption{}
\label{fig:greatPoint}
\end{subfigure}%
\caption{\textbf{Community matters.}}
\label{fig:CommunityMatters}
\end{figure}

In Fig.~\ref{fig:communitiesStickTogether}, there are 3 communities
where each community consists of 4 people (N = 12) and the
probability of interaction can be at most $u_b = 0.01$.
In the process of making the adjacency matrix, communities 
start with an interaction probability of 1/3. Hence, $u_b = 0.01$
is fairly small. However, we can see when the time is about 130,
the middle community is merged with the community at the bottom,
and consequently, the individuality tendency is very high.
The community that began in the middle separates itself from the bottom
community and its members stick together. The integrating force, i.e. interaction rule, 
overcomes the individuality tendency and glues the
agents of middle community together.

\begin{figure}[!ht]
    \centering
      \includegraphics[width=.4\linewidth]{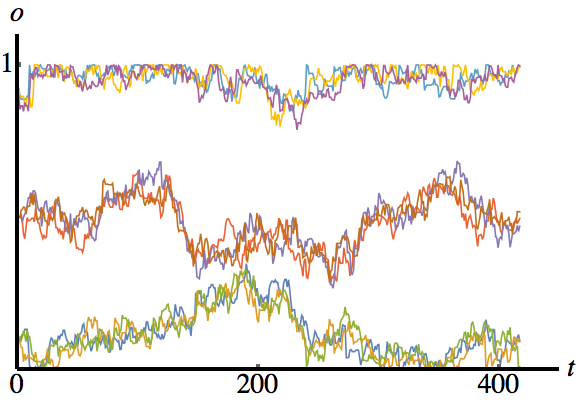}
\hspace{.4in}
\tiny
    \begin{tabular}[b]{l c c c}
       \multicolumn{2}{ c }{\tiny $ u_b=0.0025,\: \:  s = 0.00361$ }  \\ \hline
       \tiny  $ min \: \sigma_i(t)$    & \tiny  0.0092  \\ \hline
       \tiny  $ max \: \sigma_i(t)$   & \tiny 0.0199  \\ \hline
       \tiny  $ \bar{ \sigma}_i(t)$    & \tiny  0.0143  \\ \hline
       \tiny  $ \sigma(\sigma_i(t))$ & \tiny $5.755 \times 10^{-6}$  \\ \hline
    \vspace{.3in}
    \end{tabular}
    \captionlistentry[table]{A table beside a figure}
    \captionsetup{labelformat=andtable}
    \caption{\textbf{Statistics about individuality tendency.}}
     \label{fig:interesting}
  \end{figure}
\section{Identity loss is inevitable}

\begin{figure}[httb!]
\centering
\begin{subfigure}{.4\textwidth}
\includegraphics[width=1\linewidth]{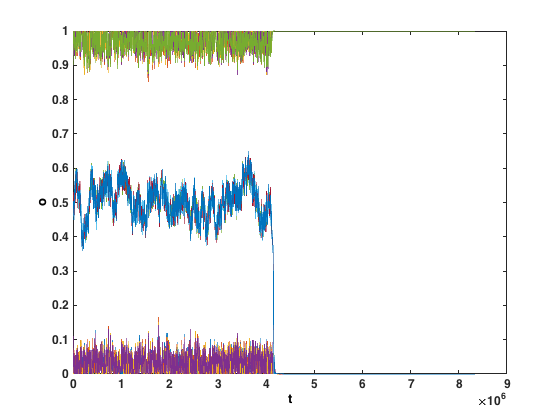}
\caption{Deterministic adjacency with $i_p = 10^{-4}, s=0$.}
\label{fig:deterministicDoomed}
\end{subfigure}%
\centering
\hspace{.01in}
\begin{subfigure}{.4\textwidth}
\includegraphics[width=1\linewidth]{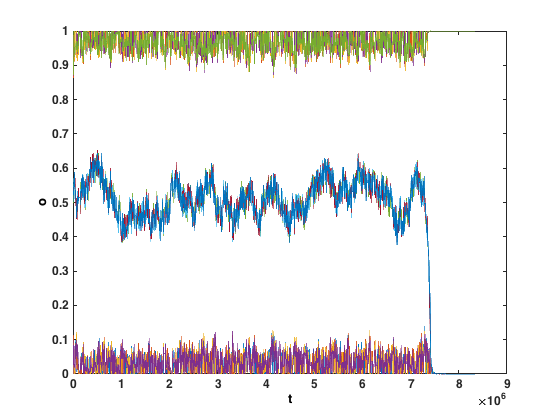}
  \caption{Random adjacency with $u_p=10^{-4}, s=0$.}
\label{fig:RandomDoomed}
\end{subfigure}%
\caption{\textbf{Identity loss is inevitable.}}
\label{fig:Doomed}
\end{figure}

With no external force countering pressure from other
communities, the slightest probability of interaction between social groups
causes loss of identity. Figure~\ref{fig:Doomed} shows
$n_s = 3$ social groups where each of which is consist of 
$n_p = 4$ nodes where probability of interaction between them
is very small, and there is no noise in the system. The game is run
for $10^{6}$ pairwise interactions. One experiment, Fig.~\ref{fig:deterministicDoomed},
uses the deterministic adjacency matrix with $i_p = 0.00001$ and the other,
Fig.~\ref{fig:RandomDoomed}, uses randomly generated matrix with $u_p = 0.00001$.
However, in order to reduce the stochastic error, 100 adjacency matrices
were generated and average of those is used in this experiment.


\section{Discussion and future work}

Results presented here shows that as individualization tendency 
(in the symmetric case) is increased the communities 
collapse more often and faster. 
However, results in~\cite{Mas2010} indicates
that individualization tendency can cause emergence of new clusters.
Please note that the result presented here does not 
contradict results of~\cite{Mas2010} for the following reasons: 
(a) we did not increase the individualization tendency parameter $s$ very much, 
since its effect is not the primary subject of our study, 
(b) we did not give the system a very long time so that 
individuals have the chance of forming a new subcommunity 
even for such small individualization tendency parameters.
It can be seen in Fig. (2) of ~\cite{Mas2010} that over time,
communities collapse and break again and again and also 
it depends on individualization tendency parameter $s$.
Moreover, their interacting update rule, which they refer to as
integrating forces, is different from that of ours. 

Topology of the network, also, plays an important
role in time evolution of opinions. Humans naturally 
tend to talk more frequently to those whom are more 
similar to, i.e. Homophily, and so, finding friends 
is an alive creature that has dynamics. Hence, coevolution of network for 
both discrete and continuous opinions  are studied~\cite{Kozma2008, 
Kozma2008-1, Nardini2008,Iniguez2009}. It would
be interesting to apply dynamic topologies to our model.
\bibliographystyle{apa}
\bibliography{community-arXiv}

\end{document}